\documentclass[aps,eqsecnum,twocolumn]{revtex4}
\usepackage{epsfig}
\usepackage{graphicx}
\usepackage{amsmath}
\usepackage{amssymb}
\usepackage{color}
\newcommand{\be}{\begin{equation}}
\newcommand{\ee}{\end{equation}}
\def\ba{\begin{eqnarray}}
\def\ea{\end{eqnarray}}
\def\pr{\partial}
\newcommand{\e}{\varepsilon}

\begin{document}
\title{Dynamical screening in bilayer graphene}
\author{O. V. Gamayun}
\email{gamayun@bitp.kiev.ua}
\affiliation{Bogolyubov Institute for Theoretical Physics, 14-b Metrologichna
str., Kiev 03680, Ukraine}

\begin{abstract}

We calculate 1-loop polarization in bilayer graphene in the 4-band approximation for  arbitrary values of frequency, momentum and doping.
 At low and high energy our result reduces to the polarization functions calculated in the 2-band approximation 
and in the case of single-layer graphehe, respectively.
The special cases of static screening and plasmon modes are analyzed.
\end{abstract}

\maketitle

\section{Introduction}

Graphene, a one-atom-thick layer of graphite, attracts a lot of attention of both theoreticians and
experimentalists since it's  fabrication \cite{graphene-fabrication}.
Quasiparticle excitations in graphene have a linear
dispersion at low energies and are described by the massless Dirac equation in 2+1 dimensions.
Theoretically such a behaviour was predicted long time ago \cite{gr} and its numerous consequences were experimentally checked after the discovery of graphene in laboratory.

Contrary to the case of single-layer graphene (SLG) low energy excitations of the bilayer graphene (BLG) have parabolic spectrum, 
although, the chiral form of the effective 2-band Hamiltonian persists because the sublattice pseudospin is still a relevant degree of freedom. 
This changes many electronic properties of the material (for review, see \cite{Kotov,DasSarmaReview}) compared to the case of monolayer graphene. 
However, the low energy approximation in bilayer graphene is
valid only for small doping $n<10^{12} cm^{-2}$, while experimentally doping can attain 10 times larger densities. 
For such a large doping, the 4-band model \cite{Falko} should be used instead of the low energy effective 2-band model.

In the literature, the screening effects in bilayer graphene were
mainly studied within the low energy effective 2-band model
\cite{h1,Nand} and in the presence of a magnetic field in
\cite{GGM,GGM1}. Dynamical polarization plays an
important role for finding plasmon excitations as well as for
studying a gap equation and excitonic condensates  both in single
layer \cite{3G,Gonzalez} and bilayer graphene \cite{Nand,GGM1}.
 Some attempts to obtain analytical results in the 4-band model for bilayer graphene were performed in
papers \cite{Borghi1,Borghi2,Kuz}. An exact calculation of the polarization function in
the 4-band model is interesting also from the pure theoretical
viewpoint because then we can see how the known results for the SLG
\cite{Guinea,Hwang} and the 2-band BLG \cite{DasSarma} are recovered as limiting cases.

Recently, a lot of attention is paid to investigate the properties of polarization operator in SLG \cite{Pyat,Asgari,Sh,Roldan1,Roldan2}. 
The most general expression for dynamical polarization of SLG at finite
temperature, chemical potential, constant impurity rate, quasiparticle
gap, and magnetic field is  given in Ref. \cite{GusPyat}.

In this paper, we calculate the BLG dynamical polarization in the 4-band model within the random phase approximation (RPA) for arbitrary wavevector, frequency and doping.
Our results can be considered as an extension of the results obtained in \cite{Borghi2}, although, those results were obtained in a slightly different approach. In Sec. \ref{Pol} we describe the
model used and present our main result for the polarization function. We consider in Sec. \ref{stat} the static polarization function and compare it with the corresponding SLG and 2-band BLG results. In Sec. \ref{plas} we focus on the long wavelength limit and study plasmons. Finally, we
provide the details of our calculations in Appendix A.

\section{RPA calculation}\label{Pol}

We model BLG in the Bernal stacking arrangement \cite{Falko},
where for two hexagonal lattices one sublattice of the bottom
layer is a near-neighbor of the opposite sublattice of the top layer.
In the tight-binding approximation, we have the following Hamiltonian:
\be
H = \sum\limits_{k,\sigma} \psi^{\sigma,\,+}_k H_k \psi^{\sigma}_k + \frac{1}{2} \sum\limits_{k}\sum\limits_{\alpha,\beta=1}^{2}\rho_k^{\alpha}V_{\alpha\beta}(k)\rho^{\beta}_{-k}\,.
\ee
Here $\psi^{\sigma}_k = (a^{\sigma}_{1}(k),b^{\sigma}_1(k),a^{\sigma}_2(k),b^{\sigma}_2(k))^{T}$, $a^{\alpha\sigma}(k)$ and $b^{\alpha\sigma}(k)$
 are destruction operators of the Bloch states of the two triangular sublattices on the graphene layers $\alpha=1,2$ with additional flavour index
$\sigma$ that encodes spin and valley. Further, $\rho_q^{\alpha}$ is the electron density on layer $\alpha$, $V_{11}(k)=V_{22}(k) = 2\pi e^2/(\kappa k)$ is the Coulomb interaction of electrons on the same layer, and electrons on different layers interact via $V_{12}(k)=V_{21}(k) = V_{11}(k) e^{-k d}$, where $d$ is the distance between the layers and $\kappa$ is dielectric permittivity of the substrate.
One-particle Hamiltonian has the following form:
\be\label{H}
H_k=\left(
      \begin{array}{cccc}
        0 & \xi\e(k) & 0 & t_{\perp} \\
        \xi\e^*(k) & 0 & 0 & 0 \\
        0 & 0 & 0 & \xi\e(k) \\
        t_{\perp} & 0 & \xi\e^*(k) & 0 \\
      \end{array}
    \right)\,,
\ee
where $t_{\perp} \sim 0.4 eV$ is the inter-layer hopping amplitude,
$\e(k)=\hbar v_F(k_x+ik_y)$, and vector $\mathbf{k} = (k_x,k_y)$ 
describes a deviation from the $K$ ($\xi=1$) and $K'$ ($\xi=-1$) point in the Brillouin zone \cite{Gusynin}. 
Below we will consider only $K$-valley.
The one-particle Hamiltonian can be diagonalized with the help of unitary matrix $U$. Then one obtains the following 4-band spectrum:
\ba
H_k = U^{-1}_k{\rm diag}(E_k^{+},-E^{+}_k,E^{-}_k,-E^{-}_k)U_k,\\
E^{\pm}_q = \sqrt{(\hbar v_F k)^2+t^2_{\perp}/4}\pm t_{\perp}/2\,.
\ea
In what follows, we put for simplicity $t_{\perp}=t$ and rescale all momenta $k\to k/\hbar v_f$. Then the Fermi momentum equals $k_F = \sqrt{\mu(\mu+t)}$ and charge density at zero temperature
is given by $n \approx k_F^2/t^2 10^{13}\, cm^{-2}$. The charge density $n=10^{12}\, cm^{-2}$ corresponds to $\mu/t = 0.1$, while the higher density $n=10^{13}\, cm^{-2}$ corresponds to $k_F=t$ and $\mu/t \approx 0.6$. 
Here $\mu$ is chemical potential (Fermi energy).
In the usual units, $k_F=t$ corresponds to $k_F\approx 0.06\, \AA^{-1}$.

If we denote the polarization matrix as follows
 $-i\langle \rho_{\alpha}(\omega,\mathbf{k})\rho_{\beta}(-\tilde{\omega},\mathbf{\tilde{k}})\rangle = \delta^{(3)}(k - \tilde{k})2\Pi_{\alpha\beta}(k)$,
then the interaction  potential in the RPA is given by
\be
V^{\rm eff}_{11}(k) = \frac{k-\alpha (1-e^{-2kd})\Pi_{11}}{k\epsilon(\omega,k)(k-\alpha(1-e^{-kd})(\Pi_{11}-\Pi_{12}))}\,,
\ee
\be
V^{\rm eff}_{12}(k) = \frac{ke^{-kd}+\alpha(1-e^{-2kd})\Pi_{12}}{k\epsilon(\omega,k)(k-\alpha(1-e^{-kd})(\Pi_{11}-\Pi_{12}))}\,,
\ee
\be
\epsilon(\omega,k)=1-\frac{\alpha(1+e^{-kd})}{k}(\Pi_{11}+\Pi_{12}).
\ee
Here $\alpha = e^2/(\hbar v_F \kappa)$ is the effective coupling constant in graphene and $d\approx 3 \AA$ is the distance between the graphene layers, which is relatively small, so we can set in
all exponents $d=0$ (even for the largest possible momentum $e^{-k_F d}\approx 0.85$). Then we have 
$V^{\rm eff}_{11}=V^{\rm eff}_{12}=1/k\epsilon(\omega,k),$
where dielectric permittivity equals $\epsilon(\omega,k) =1 -\alpha \Pi(\omega,k)/k$
with  $\Pi \equiv 2 (\Pi_{11}+\Pi_{12})$. The system is degenerate with respect to spin and valley degree of freedom therefore further we will consider polarization for one flavour degree of freedom: 
$\Pi \to \Pi/N_f$, $N_f=4$. Then 1-loop polarization is given by:
\be
\Pi(\omega,k) = T\sum_{n=-\infty}^{\infty} \int \frac{d^2q}{2\pi}{\rm Tr} G(i\Omega_n,q)G(i\Omega_n+i\omega_m,q+k).
\ee
Summation over Matsubara frequency can be easily done \footnote{Really $
T \sum_{n=-\infty}^{\infty}(i\Omega_n -\mu +a)^{-1}(i\Omega_n -\mu +b)^{-1} = (n_F(a)-n_F(b))/(a-b)
$ where $ n_F(x) = (1+\exp((x-\mu)/T))^{-1}$.}. Then performing the analytic continuation through the replacement $i\omega_m\to \omega+i0$, the retarded polarization function reads
\begin{widetext}
\be
\Pi(\omega,k) = \int \frac{d^2q}{2\pi}\sum\limits_{\alpha,\alpha'=\pm1}\sum\limits_{\lambda,\lambda'=1,2}\frac{n_F((-1)^{\lambda}E_{q}^{\alpha})-n_F((-1)^{\lambda'}E_{q+k}^{\alpha'})}{(-1)^{\lambda}E_{q}^{\alpha}-(-1)^{\lambda'}E_{q+k}^{\alpha'}-\omega-i0}F_{\lambda+1-\alpha,\lambda'+1-\alpha'}(q,q+k)\,,
\label{polarization-initial}
\ee
\end{widetext}
where indices $\lambda$ and $\alpha$ denote bands and $F_{ij}$ is a $4\times 4$ matrix responsible for the chiral structure. It is defined as follows:
\be
F_{ij}(q,p) ={\rm Tr}\left(Z^{-1}\Delta_{i}Z\Delta_{j}\right),\,\,\,\,\, Z = U_q^{-1}U_{p},
\ee
and $\Delta_j$ is diagonal matrix with all zero elements except unit at position $j$.
We can find
\be
F(q,p)= \left(
\begin{array}{cccc}
 U^{++} & V^{--} & V^{-+} & U^{+-} \\
 V^{--} & U^{++} & U^{+-} & V^{-+} \\
 V^{+-} & U^{-+} & U^{--} & V^{++} \\
 U^{-+} & V^{+-} & V^{++}& U^{--}
\end{array}
\right)\,,
\label{chirality-matrix}
\ee
where
\ba
U^{s\, u} = \frac{E^{(s)}_qE^{(u)}_p}{4E^{(0)}_qE^{(0)}_p}\left(1+s u\frac{q p \cos\theta_{qp}}{E^{(s)}_{q}E^{(u)}_{p}}\right)^2,\\
V^{s\, u} = \frac{E^{(s)}_qE^{(u)}_p}{4E^{(0)}_qE^{(0)}_p}\sin^2\theta_{qp}\,,
\ea
$E^{(s)}_{q} = \sqrt{q^2+t^2/4} +s t/2$, and $\theta_{qp}$ is the angle between vectors $\mathbf{p}$ and $\mathbf{q}$.
Diagonal elements of $F$ describe intraband transitions while off-diagonal are responsible for interband ones.
 At zero temperature, the Fermi functions in Eq.(\ref{polarization-initial}) reduce to simple step functions.
 Then our retarded polarization can be presented in the following form:
\be\label{p0}
\Pi(\omega,k) = 
\Pi^0(\omega,k)+\Pi^{+}(\omega,k)+\Pi^{-}(\omega,k)\,,
\ee
where
\ba\label{polzero}
\nonumber
&&\Pi^0(\omega,k) = \int \frac{d^2q}{\pi} \sum_{s=\pm}\left(
\frac{E_{q+k}^{(s)}+E_{q}^{(-s)}}{\omega^2 -(E_{q+k}^{(s)}+E_{q}^{(-s)})^2}U^{-s,\,s} \right.\\
&&\left.+\frac{E_{q+k}^{(-s)}+E_{q}^{(-s)}}{\omega^2 - (E_{q+k}^{(-s)}+E_{q}^{(-s)})^2}V^{s,s}
\right)\,,
\ea
\ba\label{polminus}
\nonumber
&&\Pi^{u}(\omega,k) = \int\limits_{E_q^{(u)}<\mu} \frac{d^2q}{\pi} \sum\limits_{s=\pm}\left(
\frac{usE_{q+k}^{(s)}-E_{q}^{(u)}}{\omega^2 -(usE_{q+k}^{(s)}-E_{q}^{(u)})^2}U^{u,\,s} \right.\\
&&\left.+
\frac{usE_{q+k}^{(-s)}-E_{q}^{(u)}}{\omega^2 - (usE_{q+k}^{(-s)}-E_{q}^{(u)})^2}V^{-u,s}
\right),\,\,\,\, u = \pm\,.
\ea
Clearly, $\Pi^0$ does not depend on chemical potential and characterizes the polarization at zero doping. It gives the main contribution to screening. The functions $\Pi^{+}$ and $\Pi^-$ incorporate the effects of doping and are mainly responsible for plasmon modes.
It is obvious that $\Pi^+$ can be evaluated immediately if $\Pi^-$ is found for arbitrary values of
$\mu$ and $t$. We have $\Pi^{+}_{\mu,t}=\theta(\mu -t)\Pi^{-}_{\mu-t,-t}$.

Let us comment on the chirality matrix (\ref{chirality-matrix}).
In the two limiting cases of weak ($t\to 0$) and strong ($t\to\infty$) couplings
when the spectrum reduces to $E_q = q$ and  $E_q=q^2/t$, respectively,
the chirality matrix $F$ is strongly simplified and depends only on one parameter $u_q$. Then the polarization function per one flavour degree of freedom
 \footnote{In the weak coupling regime, the number of flavours effectively doubles.} equals
\ba
 \nonumber &&\Pi(\omega,k) = \int\limits_{E_q>\mu} \frac{d^2q}{\pi}\frac{1-u_{q,k}}{2}\frac{E_q+E_{q+k}}{\omega^2-(E_q+E_{q+k})^2}\\
&&+\int\limits_{E_q< \mu }\frac{d^2q}{\pi} \frac{1+u_{q,k}}{2}\frac{E_{q+k}-E_{q}}{\omega^2-(E_{q}-E_{q+k})^2}\,,
\ea
where $u_{q,k} = ({\rm Tr} H_{q}H_{q+k})/(2E_qE_{q+k})$ which is equal to $\cos\theta_{q,q+k}$ and $\cos2\theta_{q,q+k}$
for weak and strong couplings, respectively. Note that at weak coupling $\Pi^{+}=\Pi^{-}$ while $\Pi^{+}=0$ at strong coupling.

In what follows we will consider an intermediate case for which $\mu<t$ (only this regime is experimentally relevant). In this case, $\Pi^{+}=0$. Calculation of $\Pi^0$ and $\Pi^-$ is
straightforward and the result can be written down in the following compact form:
\begin{widetext}
\be\label{pi}
\Pi(\omega,k)=\Pi^{0}(\omega,k) +\Pi^{-}(\omega,k) = 
%-\mu  + \frac{t\omega^2}{2(\omega^2-k^2)} 
-\frac{2 \mu + t}{2} - \frac{k^2 t}{4( k^2 - \omega^2)}
+ \frac{P_{\omega}+\overline{P_{-\omega}}}{4}
- c_{\omega}\overline{g}_{\omega}+\overline{g}_{t-\omega}+\overline{g}_{t+\omega}\,,
\ee
where
\be
P_{\omega} = G_{\omega+t} - c_{\omega}G_{\omega}+
i\frac{\mu_{\star}}{2}\sqrt{\frac{\rho^2_{\omega}-\mu_{\star}^2}{\omega^2-k^2}+i0\frac{k^2+\omega\mu_{\star}}{\omega^2-k^2}}
+\frac{k^2-\omega(t+\omega)}{2\omega}\log\frac{\rho^{-2}_{\omega}k^4 \mu ^2}{|\left(k^2-\omega ^2\right) \left(k^2-\omega  (2 t+\omega )\right)|}
\ee
$$
+\frac{Q_{-,\omega}^{\mu_{\star}}-Q_{+,-\omega-t}^{\omega-2\mu}+Q_{-,-\omega-t}^{2\mu -\omega}-Q_{-,-\omega}^{\mu_{\star}}}{2\omega}
-\frac{i \pi  \left|k^2-\omega (t+\omega)\right|}{2 \omega} \left(\theta\left[\omega^2-k^2-t^2\right]-\theta\left[\omega(\omega+2t)-k^2\right]\right)
$$
with
\be\label{PP}
c_{\omega}=\frac{3 k^4-k^2 \left(t^2+5 \omega^2\right)+2 \omega^4 }{2 \left(\omega^2-k^2\right)^{2}},\,\,\,\,\,\,
\rho_{\omega}=k\sqrt{\frac{\omega^2-k^2-t^2}{\omega^2-k^2}},\,\,\,\,\mu_{\star} =2\mu+t-\omega,\,\,\,g_{\omega}=\frac{\sqrt{k^2-\omega^2}}{2}\tan^{-1}\frac{\sqrt{k^2-\omega^2}}{t}\,,
\ee
\be
G_{\omega} = \sqrt{\omega^2-k^2}\left(
\log\left(\mu_{\star}{\rm sgn}(k^2-\omega^2) +\sqrt{k^2-\omega^2}\sqrt{\frac{\rho^2_{\omega}-\mu_{\star}^2}{\omega^2-k^2}+i0\frac{k^2+\omega\mu_{\star}}{\omega^2-k^2}}\right)+(\mu_{\star}\to t-\omega)
\right)\,,
\ee
\be\label{ppPp}
Q_{\pm,\omega}^r=\left|k^2-\omega (t+\omega)\right| \log \left(y+i \sqrt{{\rm sgn}\rho_{\omega}^2-y^2+i0\frac{k^2\pm \omega r}{\omega^2-k^2}}\right),\,\,\,\, y = \frac{\rho_{\omega}^2-r(\omega+t)}{\left|\rho_{\omega}(r-\omega-t)\right|}\,.
\ee
\end{widetext}
Here expressions $i0(...)$ are responsible for choosing the correct branch of the cuts. 
The square root and logarithm have a branch cut discontinuity in the complex plane running from $-\infty$ to 0. 
Equations (\ref{pi})-(\ref{ppPp}) are our main results. Details of the calculations as well as the expressions for the real and imaginary parts are given in Appendices \ref{0} and \ref{minus}.
In the weak coupling limit $t\to 0$ one can easily reproduce the results obtained in \cite{Guinea} up to the overall factor 2 which reflects the bilayer
nature of the system (in this case we should formally assume that $\mu>t$ and take into account $\Pi^{+}(\omega,k)$).
In the strong coupling limit $t\gg \mu,k,\omega$, in order to reproduce the results obtained in \cite{DasSarma} one should take into account terms of order $E_k = k^2/t$.

\begin{figure*}[htp]
\centering
\includegraphics[width=0.8\textwidth]{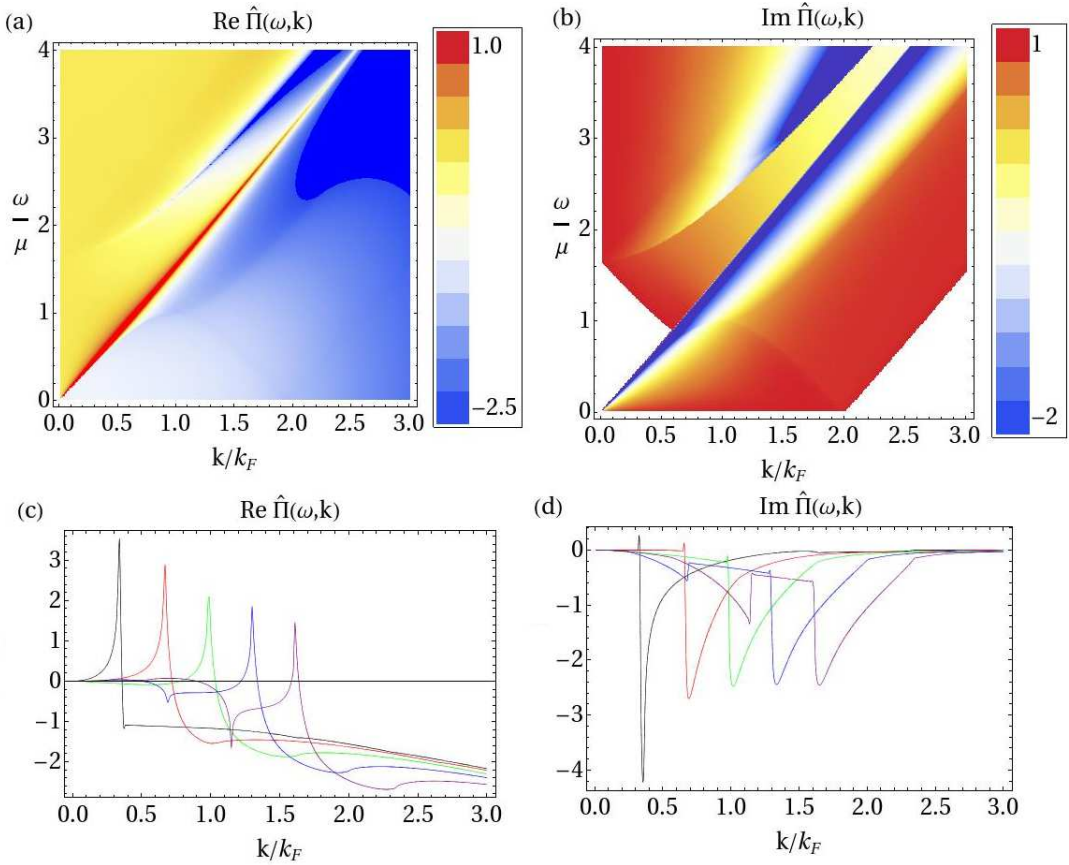}
\caption{Normalized polarization function for $\mu/t=0.6$. Panels (a) and
(b) show density plots of the real and imaginary
parts of the normalized polarization bubble defined in
Eq. (\ref{pi}), respectively. Panels (c) and (d) present constant frequency cuts for
$\omega/\mu = 0.5,\, 1.0,\, 1.5,\, 2.0,\, 2.5$.\label{FF1}}
\end{figure*}
It is convenient to normalize polarization with respect to the density of state at the Fermi level
$D(\mu) = N_f(t+2\mu)/4\pi$. So we introduce normalized polarization:
\be
\hat{\Pi}(\omega,k) \equiv -2\frac{\Pi^0(\omega,k)+\Pi^{-}(\omega,k)}{t+2\mu}.
\ee
Finally, dielectric permittivity in terms of normalized polarization is given by:
\be
\epsilon(\omega,k) = \kappa\left(1+ 2\pi\alpha D(\mu)\frac{\hat{\Pi}(\omega,k)}{k}\right),\,\,\, \alpha=  \frac{e^2}{\hbar v_F \kappa}\,.
\ee 
In Fig. \ref{FF1} we plotted $\hat{\Pi}(\omega,k)$ for $\mu/t=0.6$. One can note that the corresponding plots are very similar to those in \cite{Kotov}. The static case $\omega=0$ and the long wavelength limit $k\to 0 $ are considered in Secs. \ref{stat}, \ref{plas}.

\section{Analysis of two particular cases}

\subsection{Static screening}\label{stat}

\begin{figure}[h]
\centering
 \includegraphics[width=1.00\linewidth]{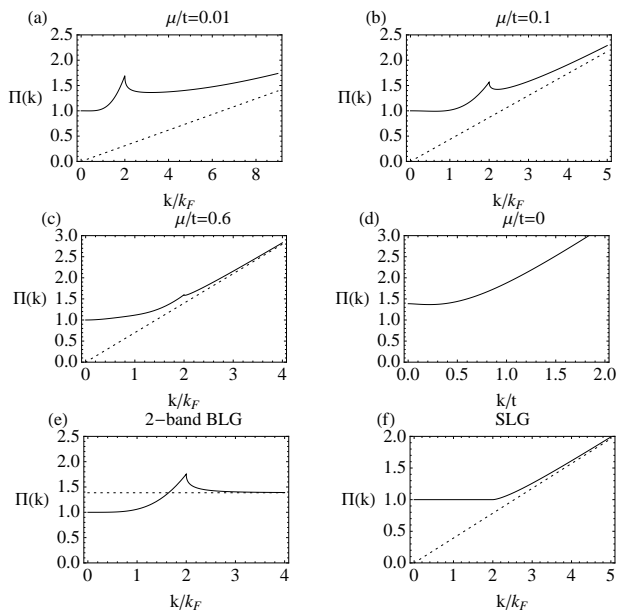}
\caption{The static polarization. Panels (a), (b) and (c) show plots of the normalized static polarization given by (\ref{zer}) at $\mu/t=0.01$, $0.1$ and $0.6$, respectively.
Dotted lines correspond to the asymptotic values $\Pi(k)=\pi k/2(t+2\mu)$.
In panel (e)  and (f) we show, respectively the static limit of the polarization function for the bilayer graphene in the 2-band approximation obtained in \cite{DasSarma}
and monolayer graphene calculated in the Dirac approximation in \cite{Guinea},\cite{2002},\cite{Kotov}. Dotted line at panel (e) corresponds to the asymptotic value $\Pi(k)=\log 4$ while
the asymptotic at panel (f) is $\Pi(k)=\pi k/8\mu$.
 \label{F1}}
\end{figure}

\begin{figure}[h]
\centering
 \includegraphics[width=1.00\linewidth]{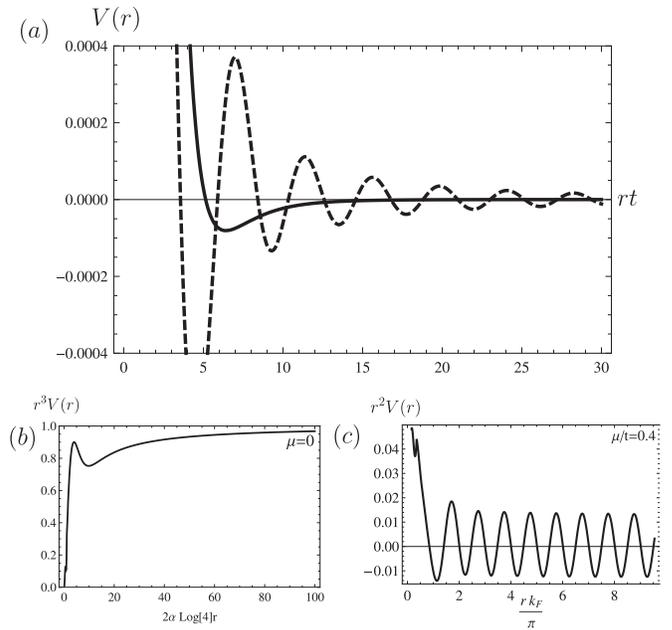}
\caption{The RPA improved Coulomb potential at finite and zero doping. At panel (a) dashed line corresponds to the potential at $\mu=0.4t$ and solid line to zero $\mu$. Panels (b) and (c) show asymptotics of the RPA improved potential at zero and finite doping, respectively.\label{F3}}
\end{figure}

The static limit  $\omega\to 0$ is relevant for screening of charged impurities. Performing some mathematical transformations we find that Eqs. (\ref{pi})-(\ref{ppPp}) imply
\begin{widetext}
\be\label{zer}
\Pi(\omega=0,k)= \frac{t}{2} \log \frac{\sqrt{k^2 + t^2}}{\mu}  - \frac{t}{4} - \mu- \frac{3 k^2 - t^2}{4 k} \tan^{-1} \frac{k}{t} -
  \sqrt{t^2 - k^2} \tanh^{-1} \frac{\sqrt{t^2 - k^2}}{t} 
\ee
$$
+\left(\sqrt{k^2 -4 \mu (t + \mu)} \left(\frac{t + 2 \mu}{2 k} + \frac{k}{2 \mu}\right) - \frac{3 k^2 - t^2}{2 k} \cos^{-1} \frac{t + 2 \mu}{\sqrt{k^2 + t^2}} -
    t \tanh^{-1} \frac{k \sqrt{k^2 - 4 \mu (t + \mu)}}{k^2 - 2 t \mu}\right) \frac{\theta[k^2 - 4 \mu (t + \mu)]}{2}
$$
$$
+
\left(\sqrt{t^2 - k^2}
   \sinh^{-1}\frac{2 \mu \sqrt{t^2 - k^2}}{k^2} - \sqrt{\frac{k^4}{4 \mu^2} - k^2 + t^2} \right)\frac{\theta[ k^4 - 4 k^2 \mu^2 + 4 t^2 \mu^2]}{2}\,.
$$
\end{widetext}
The behavior of normalized static polarization $\Pi(k)\equiv -2\hat{\Pi}(\omega=0,k)/(t+2\mu)$ and the corresponding polarizations for monolayer  \cite{Guinea}
and bilayer in the 2-band approximation \cite{DasSarma} are shown in Fig. \ref{F1}a-\ref{F1}c as functions of the normalized momentum $k/k_F$. 
We see that the polarization function calculated in the 4-band model has a discontinuity at $k=2k_F$ similar to that found in the 2-band model (see Fig. \ref{F1}e),
however, it does not go to a constant value at large momenta. Rather it grows linearly as in the case of monolayer graphene (see Fig. \ref{F1}f). For $\mu/t\to 0$, the polarization function
is similar to the polarization function in the 2-band model \cite{DasSarma} and tends to the SLG polarization function for $\mu/t\gg1$.
 Dielectric permittivity at large $k$ for bilayer graphene in the 4-band model equals $\epsilon(k) =1+\pi  \alpha N_f/4$,
whereas $\epsilon(k)=1$ in the 2-band model. Note that $\epsilon(k) =1+\pi  \alpha N_f/8$ for the SLG, therefore, we conclude
that permittivities in the BLG in the 4-band model and the SLG coincide in view of the replacement $N_f\to2N_f$ for the BLG due to doubling of the number of layers.
Since the static polarization depends only on the absolute value of momentum, the RPA improved Coulomb potential is given by the following formula:
\be
V(r) =  \int\limits_{0}^{\infty}dk \frac{kJ_0(k r)}{k + 2\pi\alpha D(\mu)\Pi(k)}\,.
\ee
At finite doping the polarization function has a discontinuity at $k=2k_F$, therefore, at large distances the potential behaves as
\be
V(r)\sim \frac{1}{r} \frac{\sin(r k_F)}{r k_F},\,\,\, r k_F\to\infty\,.
\ee
For zero doping, the discontinuity is absent and leading asymptotic is determined by the
long-wavelength behavior of the polarization function. We find
\be
V(r)\sim \frac{1}{r} \frac{1}{\left(r t\right)^{2}},\,\,\, r t\to\infty\,.
\ee
%\be
%V(r)\sim \frac{1}{r} \frac{1}{\left(2 \alpha r t \log (4) \right)^{2}},\,\,\, r t\to\infty
%\ee
The RPA improved Coulomb potential is shown in Fig. \ref{F3}.

\subsection{Plasmons}\label{plas}

The polarization function in the long wavelength limit $k\ll t$ is given by the following expression:
\ba
\nonumber
&&\Pi(\omega, k) = \frac{k^2}{2\omega^2} \left(
\mu +t+\frac{t^2 }{4 \omega}\log \frac{2 \mu +t-\omega}{2 \mu +t+\omega}+\frac{t^2}{4\omega}\log\frac{\omega-t}{t+\omega}
\right.\\
&&\left.+\frac{\omega(\omega+2t)}{4(t+\omega)}\log\frac{2\mu-\omega}{2t+\omega}-\frac{\omega(\omega-2t)}{4(t-\omega)}\log\frac{2t-\omega}{2\mu+\omega}\right)
\,.
\ea
If $\omega$ is small then
\be
\Pi(\omega, k) = \frac{k^2\mu(\mu+t)}{\omega^2(t+2\mu)}\,.
\ee
The plasmon dispersion relation is determined by the equation $\epsilon(k,\omega(k)) = 0$
which immediately gives:
\be
\omega(k) = \sqrt{k\frac{e^2 N_f}{\kappa}\frac{\mu(\mu+t)}{\mu+2t}}
\ee
\begin{figure}[h]
\centering
 \includegraphics[width=1.00\linewidth]{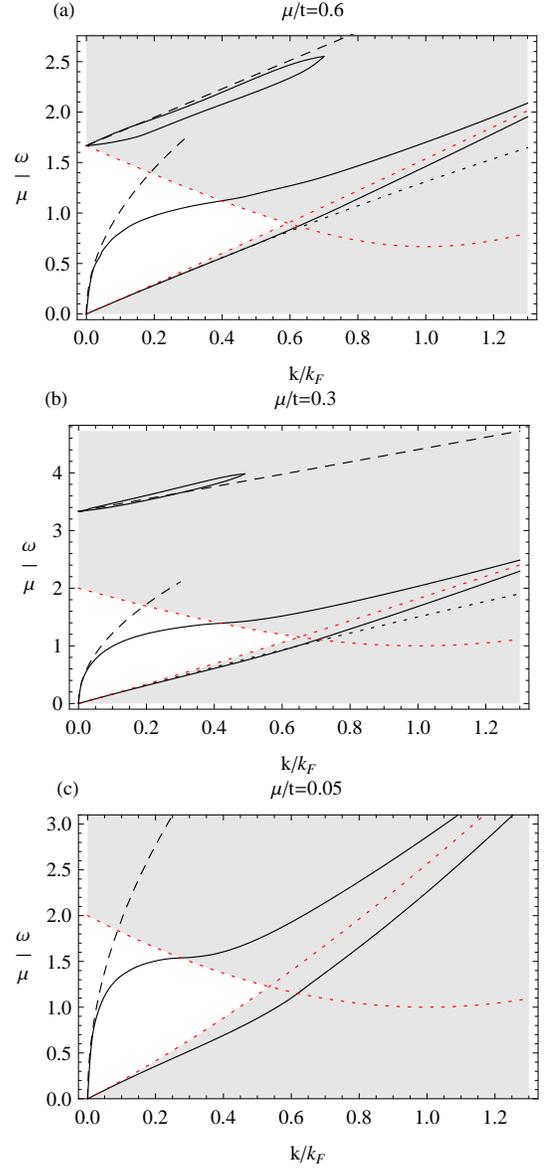}
\caption{On the panels (a), (b), (c) we present
dispersion relations (black solid lines) for plasmons in free-standing graphene at 
densities $\mu/t=0.6$, $\mu/t=0.3$ and $\mu/t=0.05$ respectively. Black dashed lines 
describe classical plasmon (\ref{pL}) and high-energy plasmon (\ref{high}). Black dotted lines 
describe additional low-energy plasmon given by (\ref{low}).
 Filled areas show domains with nonzero 
imaginary part of the polarization whose boundaries(red dotted lines) are 
given by equations (\ref{boundary}).
\label{ppp}}
\end{figure}
This is the general expression for the plasmon mode in 2D systems, which for general 
spectrum of quasiparticle excitations can be written as \cite{DasSarma}:
\be\label{pL}
\omega(k) =  \sqrt{k\frac{e^2 N_f}{2\kappa} q \frac{\pr E_q}{\pr q } \Big|_{q = q_F}}\,.
\ee
Equivalently this formula can be written as:
\be
\omega(k) = 2\pi \sqrt{k\frac{e^2 N_f}{\kappa} \frac{n}{D(\mu)}}\,,
\ee
where $n= N_f k_F^2/4\pi$ is actual two dimensional density of particle while 
$D(\mu)$ is density of states at Fermi level. In the case of the SLG $D(\mu) \sim \sqrt{n}$ so
$ \omega(k) \sim k^{1/2} n^{1/4}$.

We solve equation ${\rm Re}[\epsilon(k,\omega(k))] = 0$ numerically for free-standing graphene (i.e. $\kappa = 1$). 
Results are shown at Fig. \ref{ppp}. We see that except "classical" plasmons with low energy behaviour (\ref{pL}) we also have 
modes with linear behaviour and high-energy modes that are analogous to the $\pi$-plasmons \cite{Kat}.
The corresponding dispersion relation for small momenta are:
\ba\label{low}
&&\omega(k) = \frac{2 k (t+\mu )}{t+2 \mu }-\frac{k^2 t^2 \mu  (t+\mu )^2}{(t+2 \mu )^3}\,,\\
\label{high}
&&\omega(k) = t+\frac{e^2 N_f}{2 \kappa} k  \log \left(1+\frac{2 \mu }{t}\right).
\ea
However, contrary to the "classical" plasmons these 
modes cannot be considered as fully coherent collective
modes, because they lie in the highly damped area which corresponds to the grey filling on the plots. 
Boundaries of damped area are determined by the equation ${\rm Im}\Pi(k,\omega(k))=0$ which can be easily solved, and we obtain
\be\label{boundary}
\omega_{\pm}(k) = \sqrt{\frac{t^2}{4}+(k\pm k_F)^2}-\left|\frac{t}{2}\pm\mu\right|\,,
\ee
that describes boundary of the single particle excitations continuum (Landau damping). 
Note that contrary to the normal 2D electron gas, plasmons damp at smaller momenta due
to the interband transitions.

\section{Conclusion}

In this paper we have derived a compact analytic expression for the dynamical polarization for bilayer graphene in the 4-band model in the random phase approximation. Our results
are valid for arbitrary values of wave vector, frequency, doping and interlayer coupling. Analysing the polarization as a function of the interlayer coupling we recovered 
the expressions for the monolayer graphene polarization (weak coupling) as well as for bilayer graphene in the 2-band model (strong coupling). 
In the case where doping is smaller than the interlayer coupling we found the polarization function in the static and long-wavelength limits.
Using these results, we have obtained the RPA improved Coulomb interaction and the dispersion relation for the plasmon mode.

We put aside temperature effects and effects of the finite distance between layers, however within such formalism they 
can be easily investigated and we postpone this investigation for a separate publication. 

\begin{acknowledgments}

The author is grateful to Yu.~V.~Bezvershenko, V.~P.~Gusynin, E.~V.~Gorbar, A.~B.~Kashuba and  Y.~F.~Suprunenko for valuable discussions and useful remarks.
The author is especially grateful to Artur Slobodeniuk for collaboration at the initial stage of the work.
The work was supported partially by the SCOPES grant No. IZ73Z0 128026
of Swiss NSF, by the grant SIMTECH No. 246937 of the European FP7
program and by SFFR-RFBR grant "Application of string theory
and field theory methods to nonlinear phenomena in low dimensional
systems".

\end{acknowledgments}

\newpage
\widetext
\appendix

\section{Calculation of the polarization function}

In this appendix we present some major steps in the calculation of the normalized polarization function. All quantities are evaluated in the units of energy (Sec. \ref{Pol}).
We restrict our consideration to the case $\omega>0$ because the polarization function for negative $\omega$ can be obtained through complex conjugation.

\subsection{$\Pi^0(\omega,k)$ calculation}\label{0}

In order to calculate $\Pi^0(\omega,k)$ given by (\ref{polzero}), it is convenient to introduce the following variables:
\be
y = E_q+E_{q+k},\,\,\,\,\,\,  z= 4E_q E_{q+k}\,.
\ee
Then the measure of integration transforms as follows:
\be
\int qdq d\theta = 2\int dq^2\int\limits_0^{\pi/2} d\theta = \frac{1}{2}\int\limits_{\sqrt{k^2+t^2}}\frac{dy}{\sqrt{y^2-k^2}}\int\limits_{Q}^{y^2} \frac{zdz}{\sqrt{y^2-z}\sqrt{z-Q}}\,,
\ee
\be
Q=\frac{(y^2-k^2)^2+t^2k^2}{y^2-k^2}\,.
\ee
%So expression for the polarization operator is following:
%\be
%\Pi^0(\omega,k) =\int\limits_{\sqrt{k^2+t^2}}\frac{dy}{\pi\sqrt{y^2-k^2}}\int\limits_{Q}^{y^2} \frac{dz}{\sqrt{y^2-z}\sqrt{z-Q}}  \left(2\frac{2 y \left(y^2-k^2\right)-yz}{y^2-w^2}+G(t)+G(-t)\right)
%\ee
Performing integration over $z$, we get
\be
\Pi^0(\omega,k) =\int\limits_{\sqrt{k^2+t^2}}\frac{dy}{2}  \left(
\frac{\sqrt{y^2-k^2} (y-t)}{w^2-(y-t)^2}+\frac{\sqrt{y^2-k^2} (t+y)}{w^2-(t+y)^2}-\frac{y \left(3 k^4-k^2 \left(t^2+5 y^2\right)+2 y^4\right)}{\left(y^2-k^2\right)^{3/2} \left(w^2-y^2\right)}
\right)+\delta \Pi^0(\omega,k)\,,
\ee
where $\delta\Pi^0(\omega,k)$ is obtained by the proper change of the variables
\be
\delta\Pi^0(\omega,k) = \left(\int\limits_{\sqrt{t^2+k^2}}^{\sqrt{t^2/4+k^2}-t/2}
+\int\limits_{\sqrt{t^2+k^2}-t}^{\sqrt{t^2/4+k^2}-t/2} - \int\limits_{-t/2-\sqrt{t^2/4+k^2}}^{-\sqrt{t^2+k^2}-t}-\int\limits_{-t/2-\sqrt{t^2/4+k^2}}^{-\sqrt{t^2+k^2}}
\right)\frac{dy}{2} \frac{w^2+y t-k^2}{w^2-y^2}\,.
\ee
Now we can easily calculate the imaginary part for $\omega >0$
\be\label{IM0}
\frac{{\rm Im} \Pi^0(\omega,k)}{\pi} =\left(\frac{3 k^4-k^2 \left(t^2+5 w^2\right)+2 w^4}{4\left(w^2-k^2\right)^{3/2}}-
\frac{\left|k^2-\omega (\omega-t)\right|+\left|k^2-\omega (\omega+t)\right|}{4\omega}
\right)\theta(\omega-\sqrt{t^2+k^2})+
\ee
$$
\theta(\omega+t-\sqrt{t^2+k^2})\left(\frac{\left|k^2-\omega (\omega+t)\right|}{4\omega}-\frac{\sqrt{(\omega+t)^2-k^2}}{4}\right)+\theta(\omega-t-\sqrt{t^2+k^2})\left(\frac{\left|k^2-\omega (\omega-t)\right|}{4\omega}-\frac{\sqrt{(\omega-t)^2-k^2}}{4}\right)\,.
$$
The real part is calculated treating all divergences in the principal value sense. After some algebra we obtain
\be\label{RE0}
{\rm Re} \Pi^0(\omega,k) = \frac{k^2 t}{2(\omega^2-k^2)} -{\rm Re}\left[\frac{3 k^4-k^2 \left(t^2+5 \omega^2\right)+2 \omega^4}{2\left(k^2-\omega^2\right)^{3/2}} \tan ^{-1}\frac{\sqrt{k^2-\omega^2}}{t}\right]
\ee
$$
+\frac{\left(k^2-\omega (\omega-t)\right) }{4 \omega}\log \left|\frac{\left(k^2+(2 t-\omega) \omega\right) \left(k^2+t^2-\omega^2\right)}{\left(k^2+(t-\omega) \omega\right)^2}\right|
-\frac{k^2-\omega (t+\omega) }{4 \omega}\log \left|\frac{\left(k^2+t^2-\omega^2\right) \left(k^2-\omega (2 t+\omega)\right)}{\left(k^2-\omega (t+\omega)\right)^2}\right|
$$
$$
+{\rm Re} \left[
\frac{\sqrt{k^2-(t-\omega)^2}}{2} \tan ^{-1}\left(\frac{\sqrt{k^2-(t-\omega)^2}}{t}\right)+\frac{\sqrt{k^2-(t+\omega)^2}}{2} \tan ^{-1}\left(\frac{\sqrt{k^2-(t+\omega)^2}}{t}\right)
\right]\,.
$$
In the $t \to 0$ limit, we get
\be
\lim\limits_{t\to 0}  \Pi^0(\omega,k) = -\frac{\pi}{4}\frac{k^2}{\sqrt{k^2-\omega^2-i0}}\,,
\ee
where we performed the shift $\omega^2 \to \omega^2 +i0$ in order to reproduce the correct imaginary part.

In the large $t$ limit we must hold terms of order $k^2/t=E_k$. Then $t$ appears only as an overall factor
\be
\frac{\Pi^{0}(\omega,k)}{t/2} = \log \left|\frac{E_k^2-4\omega^2}{4E^2_k-4\omega^2}\right|+\frac{E_k}{2\omega} \log  \left|\frac{\left(E_k-\omega\right)^2}{\left(E_k+\omega\right)^2}\frac{E_k+2\omega}{E_k-2\omega}\right|+i \pi   \left(\left(1-\frac{E_k}{\omega}\right) \theta\left[\omega-E_k\right]-\left(1-\frac{E_k}{2 \omega}\right) \theta\left[2\omega-E_k\right]\right)\,.
\ee

\subsection{$\Pi^{-}(\omega,k)$ calculation} \label{minus}

In order to calculate $\Pi^{-}(\omega,k)$ given by (\ref{polminus}) we introduce new variable $r = E_q-t/2$. Then performing some algebraic manipulations we find
\be
\Pi^-(\omega,k) = \int\limits_{0}^{\mu} \frac{dr}{r}\int\limits_0^{2\pi}\frac{d\phi}{16\pi}\left(
\frac{g(\omega)}{r+\omega}+\frac{g(-\omega)}{r-\omega} - 8r-4t
\right)\,,
\ee
where
\be
g(\omega) = \frac{k^4-2 k^2 \left(2 r^2+2 r \omega+\omega (\omega-t)\right)+(2 r+\omega)^2 (t-\omega)^2}{k^2+2 k \sqrt{r (r+t)} \cos \phi +(2 r+\omega) (t-\omega)}-\frac{\left(k^2-(2 r+\omega) (2 r+t+\omega)\right)^2}{k^2+2 k \sqrt{r (r+t)} \cos \phi - \omega (2 r+t+\omega)}\,.
\ee
All divergences should be dealt with the prescription $\omega\to \omega+i0$. Then we can integrate over the angle using the following integral:
\be
\frac{1}{2\pi} \int\limits_0^{2\pi} \frac{d\phi}{a+i\epsilon 0+\cos\phi} = \frac{{\rm sgn}[a] \theta(a^2-1)}{\sqrt{a^2-1}}-i\frac{{\rm sgn}[\epsilon] \theta(1-a^2)}{\sqrt{1-a^2}}\,.
\ee
We obtain the real and imaginary parts of the polarization function:
\be
{\rm Re} \Pi^-(\omega,k) =  \int\limits_{0}^{\mu} \frac{dr}{2r} \left(\frac{{\rm Re}[g_R(\omega)]}{4(r+\omega)}+\frac{{\rm Re}[g_R(-\omega)]}{4(r-\omega)}- 2r-t
\right),\,\,\,\,
{\rm Im} \Pi^-(\omega,k) = - \int\limits_{0}^{\mu} \frac{dr}{8r} \left(\frac{{\rm Re}[g_I(\omega)]}{r+\omega}-\frac{{\rm Re}[g_I(-\omega)]}{r-\omega}\right)\,,
\ee
\be
g_R(\omega) = \sqrt{\left(k^2+(2 r+\omega) (t-\omega)\right)^2-4 k^2 r (r+t)} {\rm sgn}\left(k^2+(2 r+\omega) (t-\omega)\right)
\ee
$$
-\frac{\left(k^2-(2 r+\omega) (2 r+t+\omega)\right)^2 {\rm sgn}\left(k^2-\omega (2 r+t+\omega)\right)}{ \sqrt{\left(k^2-\omega (2 r+t+\omega)\right)^2-4 k^2 r (r+t)}}\,,
$$
\be
g_I(\omega) = \sqrt{4 k^2 r (r+t)-\left(k^2+(2 r+\omega) (t-\omega)\right)^2} {\rm sgn}\left(r-\frac{t}{2}+\omega\right)+\frac{\left(k^2-(2 r+\omega) (2 r+t+\omega)\right)^2 {\rm sgn}\left(r+\frac{t}{2}+\omega\right)}{\sqrt{4 k^2 r (r+t)-\left(k^2-\omega (2 r+t+\omega)\right)^2}}\,.
\ee

We can calculate all integrals separately keeping regularization $\epsilon$ of possible divergences at $r=0$. In order to write down the answer in a compact form, we introduce the following notation. For any given function $f(x)$, one can construct a new function 
$\widehat{f}(x)\Big|_a^b$ by the following rule \footnote{One can easily note that $\hat{f}(x)\Big|^{b}_a = \int_{a}^b dr f^{\prime}(r){\rm sgn}(r-x)$. }:
\be
\widehat{f}(x)\Big|^{b}_a \equiv {\rm sgn}(b-x)(f(b)-f(r))-{\rm sgn}(a-x)(f(a)-f(x))
\ee
Then one can present the polarization in the following form:
\be\label{regmin}
\Pi_{\epsilon}^-(\omega,k)  =  -\mu -\frac{t}{2}\log\frac{2\mu}{\epsilon}+ \frac{{\rm Re}(R_{\omega}+R_{-\omega})+ i{\rm Re}(I_{\omega}-I_{-\omega})}{2} -i \pi \frac{|k^2-(t+\omega)\omega|}{4\omega}\theta(\mu -\omega)(\theta(\rho_{t+\omega}^2)-\theta(-\rho_{\omega}^2))\,,
\ee
where $R_{\omega}=R^{\epsilon}_{\omega}+\tilde{R}_{\omega}$, $I_{\omega}=I^{\epsilon}_{\omega}+\tilde{I}_{\omega}$, and
\be
R^{\epsilon}_{\omega} = i\frac{|k^2+(t-\omega)\omega|}{2\omega} \left(
\widehat{\tilde{f}_{t-\omega}^{\omega}}\left(\frac{-k^2}{\omega}\right)\Big|^{2\tilde{\mu}-\omega}_{t-\omega+\epsilon}
+\widehat{\tilde{f}_{\omega}^{t-\omega}}\left(\frac{k^2}{\omega-t}\right)\Big|^{2\mu+\omega}_{\omega+\epsilon}
-\widehat{\tilde{f}_{t-\omega}^{\omega}}\left(\frac{k^2}{\omega}\right)\Big|^{2\tilde{\mu}+\omega}_{t+\omega}
-\widehat{\tilde{f}_{-\omega}^{t-\omega}}\left(\frac{k^2}{\omega-t}\right)\Big|^{2\mu+\omega}_{\omega}
\right)\,,
\ee
\be
\tilde{R}_{\omega}=\widehat{\tilde{v}_{\omega}}\left(\frac{-k^2}{\omega}\right)\Big|^{2\tilde{\mu}-\omega}_{t-\omega} - \frac{3k^4 -k^2t^2-5k^2\omega^2+2\omega^4}{2(\omega^2-k^2)}\widehat{\tilde{u}_{\omega}}\left(\frac{-k^2}{\omega}\right)\Big|^{2\tilde{\mu}-\omega}_{t-\omega}
+((\omega-t)^2-k^2)\widehat{\tilde{u}_{t-\omega}}\left(\frac{k^2}{\omega-t}\right)\Big|^{2\mu+\omega}_{\omega}\,,
\ee
\be
I^{\epsilon}_{\omega} = \frac{|k^2+(t-\omega)\omega|}{2\omega} \left(
\widehat{f^{\omega}_{t-\omega}}(-\omega)\Big|^{2\tilde{\mu}+\omega}_{t+\omega} 
-\widehat{f^{t-\omega}_{-\omega}}(t-\omega)\Big|^{2\mu+\omega}_{\omega}
-\widehat{f^{\omega}_{t-\omega}}(\omega)\Big|^{2\tilde{\mu}-\omega}_{t-\omega+\epsilon} 
+\widehat{f^{t-\omega}_{\omega}}(t-\omega)\Big|^{2\mu+\omega}_{\omega+\epsilon}
\right)\,,
\ee
\be
\tilde{I}_{\omega}=\widehat{v_{\omega}}(\omega)\Big|^{2\tilde{\mu}-\omega}_{t-\omega} - \frac{3k^4 -k^2t^2-5k^2\omega^2+2\omega^4}{2(\omega^2-k^2)}\widehat{u_{\omega}}(\omega)\Big|^{2\tilde{\mu}-\omega}_{t-\omega}
-((\omega-t)^2-k^2)\widehat{u_{t-\omega}}(t-\omega)\Big|^{2\mu+\omega}_{\omega}\,.
\ee
Here
\be
\rho_{\omega}=\sqrt{k^2\frac{\omega^2-k^2-t^2}{\omega^2-k^2}},\,\,\, \tilde{\mu}= \mu+t/2,\,\,\, f^{\omega}_{\Omega}(r)=\tan^{-1}\left(\frac{e^{i\sin ^{-1}(r/\rho_{\omega})}-i \Omega/\rho_{\omega}}{\sqrt{\Omega^2/\rho^2_{\omega}-1}}\right)
,\,\,\, \tilde{f}^{\omega}_{\Omega}(r)={\rm sgn}(\omega(\omega^2-k^2-t^2))f^{\omega}_{\Omega}(r)
\ee
and
\be
v_{\omega}(x) =\frac{x\sqrt{\rho_{\omega}^2-x^2}}{4\sqrt{k^2-\omega^2}},\,\,\,\,\,\, \tilde{v}_{\omega}(x)={\rm sgn}(\omega(k^2-\omega^2))v_{\omega}(x),\,\,\,
u_{\omega}(x) =\frac{\sin^{-1}(x/\rho_{\omega})}{4\sqrt{k^2-\omega^2}},\,\,\,\,\,\, \tilde{u}_{\omega}(x)={\rm sgn}(\omega(k^2-\omega^2))u_{\omega}(x)\,.
\ee
Expression (\ref{regmin}) should be understood in the limit $\epsilon \to 0$. Taking this limit explicitly, we find
\be\label{mu}
\Pi^-(\omega,k)  =  -\mu + \frac{{\rm Re}(R^{\rm reg}_{\omega}+R^{\rm reg}_{-\omega}+\tilde{R}_{\omega}+\tilde{R}_{-\omega})}{2}+ i\frac{{\rm Re}(I^{\rm reg}_{\omega}-I^{\rm reg}_{-\omega}+\tilde{I}_{\omega}-\tilde{I}_{-\omega})}{2} 
\ee
$$
-i \pi \frac{|k^2-(t+\omega)\omega|}{4\omega}\left(\theta(\mu -\omega)(\theta(\rho_{t+\omega}^2-\theta(-\rho_{\omega}^2))-\frac{\pi}{2}\theta(\omega^2 - k^2 -t^2))
\right)
- i\pi \frac{|k^2-(t-\omega)\omega|}{4\omega}\theta((\omega-t)^2-\omega^2-k^2)\,,
$$
where 
\be
R^{\rm reg}_{\omega} = i \frac{|k^2+(t-\omega)\omega|}{2\omega}\left(
 \tilde{G}^{2\tilde{\mu}-\omega, -k^2/\omega}_{\omega,t-\omega} -
\tilde{G}^{2\tilde{\mu}+\omega, k^2/\omega}_{\omega,t-\omega} + 
\tilde{G}^{t+\omega, k^2/\omega}_{\omega,t-\omega}
+
\tilde{G}^{2\mu+\omega,k^2/(\omega-t)}_{t-\omega,\omega}-
\tilde{G}^{2\mu+\omega,k^2/(\omega-t)}_{t-\omega,-\omega}+
\tilde{G}^{\omega,k^2/(\omega-t)}_{t-\omega,-\omega}
\right)\,,
\ee
$$
+\frac{k^2+(t-\omega)\omega}{2\omega} \left(
i{\rm sgn}(\omega)\tilde{f}^{\omega}_{t-\omega}\left(\frac{-k^2}{\omega}\right)-\frac{1}{2}\log\mu\frac{\sqrt{\rho_{\omega}^2-(t-\omega)^2}+i(t-\omega)}{\rho_{\omega}^2-(t-\omega)^2}+(\omega\to t-\omega)
\right)\,,
$$
\be
I^{\rm reg}_{\omega} = \frac{|k^2+(t-\omega)\omega|}{2\omega}\left(
G^{2\tilde{\mu}+\omega,-\omega}_{\omega,t-\omega}-G^{t+\omega,-\omega}_{\omega,t-\omega}-G^{2\mu+\omega,t-\omega}_{t-\omega,-\omega}
+G^{\omega,t-\omega}_{t-\omega,-\omega}-G^{2\tilde{\mu}-\omega,\omega}_{\omega,t-\omega}+G^{2\mu+\omega,t-\omega}_{t-\omega,\omega}
\right)
\ee
$$
+\frac{|k^2+(t-\omega)\omega|}{2\omega}\left(\theta(k^2-\omega^2)\cos^{-1}\left(\frac{t-\omega}{\rho_{\omega}}\right)-\frac{\pi}{2}\theta(-\rho_{\omega}^2)-f^{\omega}_{t-\omega}(\omega) +(\omega \to t-\omega)\right)\,,
$$
and
\be
G^{a,b}_{\omega,\Omega} ={\rm sgn} (b-a)\left(f_{\Omega}^{\omega}(b)-f_{\Omega}^{\omega}(a)\right),\,\,\,\,\,\,\,\,
\tilde{G}^{a,b}_{\omega,\Omega} ={\rm sgn} (b-a)\left(\tilde{f}_{\Omega}^{\omega}(b)-\tilde{f}_{\Omega}^{\omega}(a)\right)\,.
\ee
In the weak coupling limit $t\to 0$, we have the following expression:
\be
\Pi^-(\omega,k)  =  -\mu + \frac{{\rm Re}(R^{t=0}_{\omega}+R^{t=0}_{-\omega})+ i{\rm Re}(I^{t=0}_{\omega}-I^{t=0}_{-\omega})}{2}\,, 
\ee
\be
R^{t=0}_{\omega} = \frac{{\rm sgn}\left(\left(k^2-\omega^2\right) \left(k^2-\omega^2+2 \omega \mu \right)\right) \left((2 \mu -\omega) \sqrt{k^2-(2 \mu -\omega)^2}+k^2 \left(\sin ^{-1}\left(k/\omega\right)-\sin ^{-1}\left((\omega-2 \mu)/k\right)\right)\right)}{4 \sqrt{k^2-\omega^2}}\,,
\ee
\be
I^{t=0}_{\omega} = {\rm sgn}(\mu -\omega) \left(\frac{(2 \mu -\omega) \sqrt{k^2-(2 \mu -\omega)^2}}{4 \sqrt{k^2-\omega^2}}-\frac{k^2 \left(\sin ^{-1}\left((\omega-2 \mu)/k\right)+\sin ^{-1}\left(\omega/k\right)\right)}{4 \sqrt{k^2-\omega^2}}-\frac{\omega}{4}\right)\,.
\ee
We can unite real and imaginary part in one expression:
\be
\Pi^{-}(\omega,k) = -\mu  - \frac{t(k^2-2\omega^2)}{2(\omega^2-k^2)} + \frac{P_{\omega}+\overline{P_{-\omega}}}{4}\,,
\ee
where
\be
P_{\omega} = G_{\omega+t} - \frac{3 k^4-k^2 \left(t^2+5 \omega^2\right)+2 \omega^4 }{2 \left(\omega^2-k^2\right)^{2}}G_{\omega}+
i\frac{\mu_{\star}}{2}\sqrt{\frac{\rho^2_{\omega}-\mu_{\star}^2}{\omega^2-k^2}+i0\frac{k^2+\omega\mu_{\star}}{\omega^2-k^2}}
+\frac{Q_{-,\omega}^{\mu_{\star}}-Q_{+,-\omega-t}^{\omega-2\mu}+Q_{-,-\omega-t}^{2\mu -\omega}-Q_{-,-\omega}^{\mu_{\star}}}{2\omega}
\ee
$$
+\frac{k^2-\omega(t+\omega)}{2\omega}\log\frac{\rho_{\omega}^2\mu^2|(\omega^2-k^2)(\omega(\omega+2t)-k^2)|}{(k^2-\omega(\omega+t))^4}
+\frac{i \pi  \left|k^2-\omega (t+\omega)\right|}{2 \omega} \left(\theta\left[\omega^2-k^2-t^2\right]-\theta\left[\omega(\omega+2t)-k^2\right]\right),
$$
and the functions $G_{\omega}$ and $Q^{r}_{\pm,\omega}$ are determined in Eqs.(\ref{PP})-(\ref{ppPp}).

\end{document}